\documentclass[aps,prl,twocolumn,superscriptaddress]{revtex4}

\usepackage{mathrsfs,amssymb,graphics,subfigure,threeparttable}
\usepackage{grap hicx}
\usepackage{amssymb}
\usepackage{enumerate}
\usepackage{amsmath}
\usepackage{fancyhdr}
\usepackage{bm}
\usepackage[squaren]{SIunits}
\usepackage{hyperref}

\usepackage{xcolor}

\begin{document}


\title{\textit{In Situ} Generation of High-Energy Spin-Polarized Electrons in a Beam-Driven Plasma Wakefield Accelerator}


\author{Zan Nie}
\email[]{znie@ucla.edu}
\affiliation{Department of Electrical and Computer Engineering, University of California Los Angeles, Los Angeles, California 90095, USA}
\author{Fei Li}
\email[]{lifei11@ucla.edu}
\affiliation{Department of Electrical and Computer Engineering, University of California Los Angeles, Los Angeles, California 90095, USA}

\author{Felipe Morales}
\author{Serguei Patchkovskii}
\author{Olga Smirnova}
\affiliation{Max Born Institute, Max-Born-Str. 2A, D-12489 Berlin, Germany}

\author{Weiming An}
\affiliation{Department of Astronomy, Beijing Normal University, Beijing 100875, China}

\author{Noa Nambu}
\author{Daniel Matteo}
\author{Kenneth A. Marsh}
\affiliation{Department of Electrical and Computer Engineering, University of California Los Angeles, Los Angeles, California 90095, USA}

\author{Frank Tsung}
\affiliation{Department of Physics and Astronomy, University of California Los Angeles, Los Angeles, California 90095, USA}
\author{Warren B. Mori}
\affiliation{Department of Electrical and Computer Engineering, University of California Los Angeles, Los Angeles, California 90095, USA}
\affiliation{Department of Physics and Astronomy, University of California Los Angeles, Los Angeles, California 90095, USA}

\author{Chan Joshi}
\email[]{cjoshi@ucla.edu}
\affiliation{Department of Electrical and Computer Engineering, University of California Los Angeles, Los Angeles, California 90095, USA}

\date{\today}

[Corrected version]

\begin{abstract}

\textit{In situ} generation of a high-energy, high-current, spin-polarized electron beam is an outstanding scientific challenge to the development of plasma-based accelerators for high-energy colliders. In this Letter we show how such a spin-polarized relativistic beam can be produced by ionization injection of electrons of certain atoms with a circularly polarized laser field into a beam-driven plasma wakefield accelerator, providing a much desired one-step solution to this challenge. Using time-dependent Schrödinger equation (TDSE) simulations, we show the propensity rule of spin-dependent ionization of xenon atoms does not flip as a function of photoelectron energy compared with the non-adiabatic tunneling regime, leading to high total spin polarization. Furthermore, three-dimensional particle-in-cell (PIC) simulations are incorporated with TDSE simulations, providing start-to-end simulations of spin-dependent strong-field ionization of xenon atoms and subsequent trapping, acceleration, and preservation of electron spin-polarization in lithium plasma. We show the generation of a high-current (0.8\,kA), ultra-low-normalized-emittance ($\sim$\,37\,nm), and high-energy (2.7\,GeV) electron beam within just 11\,cm distance, with up to $\sim$\,31\% net spin polarization. Higher current, energy, and  net spin-polarization beams are possible by optimizing this concept, thus solving a long-standing problem facing the development of plasma accelerators.

\end{abstract}


\maketitle

In high-energy lepton colliders, collisions between spin-polarized electron and positron beams are preferred \cite{barish2013international}. Spin-polarized relativistic particles are chiral and therefore ideally suited for selectively enhancing or suppressing specific reaction channels and thereby better characterizing the quantum numbers and chiral couplings of the new particles. To enable science at the ever-increasing energy frontier of elementary particle physics while simultaneously shrinking the size and cost of future colliders, development of advanced accelerator technologies is considered essential. While plasma-based accelerator (PBA) schemes have made impressive progress in the past three decades, a concept for \textit{in situ} generation of spin-polarized beams has thus far proven elusive. The most common spin-polarized electron sources are based on photoemission from a Gallium Arsenide (GaAs) cathode \cite{pierce1976photoemission}. Spin-polarized positron beams may be obtained from pair production by polarized bremsstrahlung photons, the latter produced by passing a spin-polarized relativistic electron beam through a high-Z target \cite{abbott2016production}. Unfortunately, none of the above methods can generate ultra-short (few microns long) and precisely (fs) synchronized spin-polarized electron beams necessary for injection into PBAs. 

The only previous proposal for producing spin-polarized electron beams from PBA \cite{vieira2011polarized,wen2019polarized,wu2019polarized1,wu2019polarized} involves injecting spin-polarized electrons into a wake excited by a moderate intensity laser pulse or a moderate charged electron beam in a density down-ramp. However, this proposal is a two-step scheme. The first step requires the generation of spin-polarized electrons outside of the PBA set-up by employing a complicated combination (involving multiple lasers) of molecular alignment, photodissociation and photoionization of hydrogen halides \cite{sofikitis2017highly,sofikitis2018ultrahigh}. Even though the spin polarization of the hydrogen atoms can be high, the overall net spin polarization of electrons ionized from both hydrogen and halide atoms is expected to be low \cite{wen2019polarized}. The second step involves the injection of these spin-polarized electrons crossing the strong electromagnetic fields of the plasma wake. To avoid severe spin depolarization due to these strong electromagnetic fields, the wakefield should be moderately strong, which limits both the accelerating gradient and charge of the injected electrons.

In the one-step solution we propose here, the generation and subsequent acceleration of spin-polarized electrons is integrated within the wake itself. Using a combination of TDSE \cite{patchkovskii2016simple,manolopoulos2002derivation,morales2016isurf}  and 3D-PIC \cite{fonseca2002osiris,fonseca2008one,li2020quasi} simulations, we show that spin-polarized electrons can be produced \textit{in situ} directly inside a beam-driven plasma wakefield accelerator and rapidly accelerated to multi GeV energies by the wakefield without significant depolarization. Electrons are injected and simultaneously spin-polarized via ionization of the outermost p-orbital of a selected noble gas (no need for pre-alignment) using a circularly polarized laser \cite{barth2013spin}. The mitigation of depolarization is another benefit of laser-induced ionization injection \cite{oz2007ionization,pak2010injection}: the electrons can be produced inside the wake close to the wake axis, where the transverse magnetic and electric fields of the wake are near zero  \cite{lu2006nonlinear}, minimizing both the beam emittance and depolarization due to spin precession. A third advantage of our scheme is that the wake can be in the highly nonlinear or bubble regime where electrons are rapidly accelerated to $c$ minimizing the emittance growth and accelerating the electrons at higher gradients. 

The proposed experimental layout of our scheme is shown in Supplementary Material \cite{SM}. A relativistic drive electron beam traverses a column of gas containing a mixture of lithium (Li) and xenon (Xe) atoms. The ionization potentials of the $2s$ electron of Li atoms and the outermost $5p^6$ electron of Xe atoms are 5.4 eV and 12.13 eV, respectively. The electron beam fully ionizes Li atoms and produces the wake while keeping Xe atoms unionized. If the driving electron beam is ultra-relativistic ($\gamma \gg 1$) and sufficiently dense ($n_b>n_p$, $k_p \sigma_{r,z}<1$), the $2s$ electrons of the Li atoms are ionized during the risetime of the beam current and blown out by the transverse electric field of the beam to form a bubble-like wake cavity \cite{lu2006nonlinear,litos2014high} that contains only the Li ions and the neutral Xe atoms. Now an appropriately delayed circularly polarized ultra-short laser pulse copropagating with the electron beam is focused at the entrance of the Li plasma to strong-field ionize the $5p^6$ electron of the Xe atoms, producing spin-polarized electron beam close to the center (both transversely and longitudinally) of the first bucket of the wake. The injected electrons are subsequently trapped by the wake potential and accelerated to $\sim$\,2.7\,GeV energy in $\sim$\,11\,cm without significant depolarization.

\begin{figure}[tp]
\includegraphics[width=0.5\textwidth]{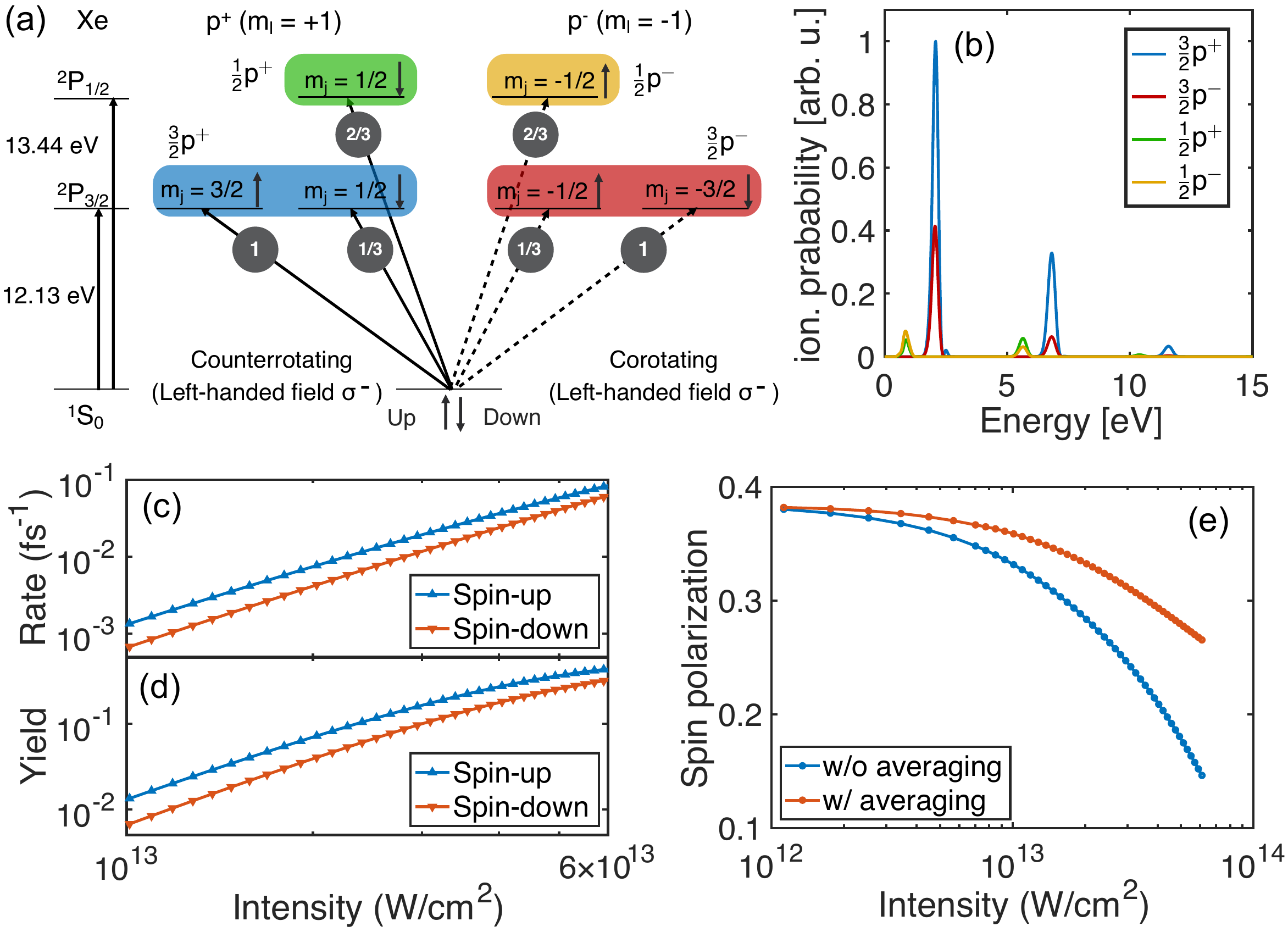}
\caption{\label{fig1} 
(a) Schematic of spin-dependent photoionization showing possible ionization pathways from Xe to Xe$^+$. (b) TDSE simulation results of the multi-photon ionization photoelectron spectra for the final ionic state, Xe$^+(^2\text{P}_{3/2})$ or Xe$^+(^2\text{P}_{1/2})$, the energy and the initial quantum number $m_l=\pm 1$ of the photoelectron, for 10\,fs (FWHM), $\lambda=260$\,nm laser pulse with peak intensity $I=2.5\times 10^{13}\,\text{W/cm}^2$. (c,d) Log-log plot of the simulated ionization rates and yields of spin-up and spin-down electrons as a function of laser peak intensity of a 260\,nm, 10\,fs (FWHM), circularly polarized laser. (e) Spin polarization as a function of peak laser intensity without and with focal-volume averaging.
}
\end{figure}

It is known that strong field ionization rate of a fixed orbital in circularly polarized fields depends on the sense of electron rotation (i.e. the magnetic quantum number $m_l$) in the initial state \cite{Popruzhenko2008, barth2011nonadiabatic, barth2013nonadiabatic}. Based on this phenomenon and spin-orbit interaction in the ionic core, spin-polarized electrons can be produced by strong-field ionization  \cite{barth2013spin}. Here we use Xe atoms as an example, but there are many other possibilities. Xe has six $p$-electrons in its outermost shell, with $m_l\equiv l_z=0,\pm 1$. Strong-field ionization from the $p^0$ orbital ($m_l=0$) in circularly polarized laser fields is negligible in the strong–field regime \cite{barth2011nonadiabatic,barth2013nonadiabatic}. Here, we choose left-handed ($\sigma^-$) circularly polarized pulses both in our analysis and later calculations. Consider first ionization from the $p^+$ orbital (counter-rotating with the laser field) into the two lowest states of Xe$^+$, $^2\text{P}_{3/2}$ and $^2\text{P}_{1/2}$, see the left half of the ionization pathways in Fig.\,\ref{fig1}(a). Removal of a spin-up $p^+$ electron ($s_z=1/2$, $l_z=1$) would create a hole with $j_z=+3/2$ and could only generate the ion in the state $^2\text{P}_{3/2}$. Removal of a spin-down $p^+$ electron ($s_z=-1/2$, $l_z=1$) would create a hole with $j_z=+1/2$ and can generate the ion both in the $^2\text{P}_{3/2}$ and $^2\text{P}_{1/2}$ states, with the Clebsch-Gordan coefficients squared splitting the two pathways as 1/3 for $^2\text{P}_{3/2}$ and 2/3 for $^2\text{P}_{1/2}$. Repeating the same analysis for the $p^-$ electron (right half of ionization pathways in Fig.\,\ref{fig1}(a)), one obtains the following expressions for the ionization rates $W_\uparrow$ and $W_\downarrow$ of spin-up and spin-down electrons \cite{barth2013spin}:
\begin{align}
&W_{\uparrow}=W_{\frac{3}{2}p^+}+\frac{2}{3}\,W_{\frac{1}{2}p^-}+\frac{1}{3}\,W_{\frac{3}{2}p^-}  \label{eq1} \\
&W_{\downarrow}=W_{\frac{3}{2}p^-}+\frac{2}{3}\,W_{\frac{1}{2}p^+}+\frac{1}{3}\,W_{\frac{3}{2}p^+} \label{eq2}
\end{align}
where $W_{\frac{3}{2}p^+}$, $W_{\frac{3}{2}p^-}$, $W_{\frac{1}{2}p^+}$, and $W_{\frac{1}{2}p^-}$ denote ionization rates of a $p^+$ electron into the $^2\text{P}_{3/2}$ state, a $p^-$ electron into the $^2\text{P}_{3/2}$ state, a $p^+$ electron into the $^2\text{P}_{1/2}$ state, and a $p^-$ electron into the $^2\text{P}_{1/2}$ state, respectively. Net spin polarization arises under two conditions: (i) either $p^+$ ionization dominates $p^-$ or vice versa and (ii) one of the two ionic states is more likely to be populated.

In the adiabatic tunneling regime of strong-field ionization (Keldysh parameter \cite{keldysh1965ionization} $\gamma_\text{K}\ll 1$), the ionization rates of $p^+$ and $p^-$ electrons are the same and ionization is not spin-selective. In the non-adiabatic tunneling regime ($\gamma_\text{K}\sim\,1$) \cite{ivanov2005anatomy}, counter-rotating electrons are more likely to be ionized \cite{barth2011nonadiabatic, barth2013nonadiabatic, herath2012strong, eckart2018ultrafast}, and the population of Xe$^+(^2\text{P}_{1/2})$ is suppressed due to its higher ionization potential ( I$_\text{p}\,(^2\text{P}_{1/2}) = 13.44$\,eV compared to I$_\text{p}\,(^2\text{P}_{3/2})= 12.13$\,eV), satisfying both conditions for generating spin-polarized electrons. Both the $m_l$-dependent ionization rates and the resulting spin polarization have been experimentally verified \cite{hartung2016electron, herath2012strong, eckart2018ultrafast, trabert2018spin, liu2018energy}. However, the observed spin polarization generated by ionization of Xe at 800 nm and 400 nm changes sign both between the two ionization channels and across the photoelectron spectrum \cite{barth2013spin, hartung2016electron, trabert2018spin, liu2018energy}, reducing the net spin polarization upon integrating over all photoelectron energies and both ionic states. 

Theory and simulations show that propensity rules for ionization can be reversed in the multi-photon regime ($\gamma_\text{K}\gg 1$) \cite{Bauer2014,Zhu2016,Xu2020}. From our TDSE simulations, ionization of Xe by a $\lambda$=260 nm left-handed ($\sigma^-$) circularly polarized laser is strongly dominated by the removal of a $p^+$ electron (counter-rotating with the laser field) at all laser intensities, until saturation, and for all photoelectron energies, with ionization into Xe$^+(^2\text{P}_{1/2})$ strongly suppressed (Fig.\,\ref{fig1}(b)), which leads to high total spin-polarization. We have performed simulations for a range of intensities from $3.5\times 10^{10}\,\text{W/cm}^2$ to $6.3\times 10^{13}\,\text{W/cm}^2$, by solving the TDSE for each intensity for four ionization pathways: $\frac{3}{2} p^+$, $\frac{1}{2} p^+$, $\frac{3}{2} p^-$, and $\frac{1}{2} p^-$, and calculated the corresponding spin-up and spin-down electron ionization rates and yields (Fig.\,\ref{fig1}(c,d)) according to Eq.\,(\ref{eq1})(\ref{eq2}). The net spin-polarization with integration over temporal and spatial intensity distribution, all photoelectron energies, and final ionic states (see Supplementary Material of Ref.\,\cite{zimmermann2017unified}) is shown in Fig.\,\ref{fig1}(e). For the laser intensity we used in the following PIC simulations ($I=2.5\times 10^{13}\,\text{W/cm}^2$), the net spin-polarization reached 32\% after focal-volume averaging.

\begin{figure}[tp]
\includegraphics[width=0.5\textwidth]{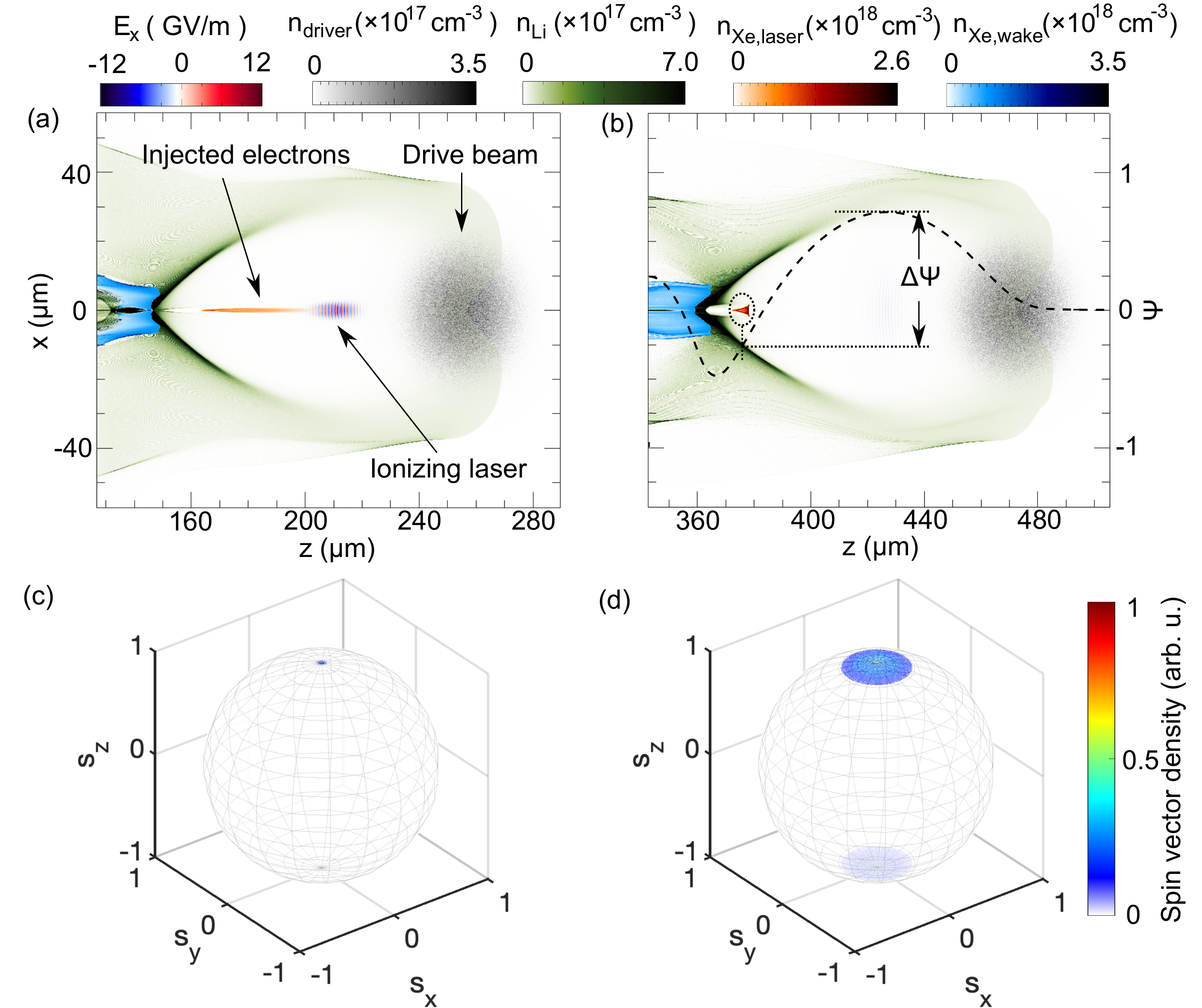}
\caption{\label{fig2} 
(a),(b) Two snapshots show the charge density distribution of driving electron beam (grey), beam ionized Li electrons (green), laser ionized Xe electrons (brown), and wakefield ionized Xe electrons (blue) at (a) $z=210\,\mu$m (end of ionization injection) and (b) $z=425\,\mu$m (after being trapped). The dashed lines in (b) show the on-axis wake pseudo potential. The wakefield ionized Xe electrons (blue) are only generated at the tail of the bubble and cannot be trapped by the wake. (c),(d) The spin vector density distribution of Xe electrons ionized by the UV laser at the same moment of (a) and (b).
}
\end{figure}

We have incorporated the spin-dependent ionization results into our wakefield acceleration simulations. By tightly focusing a 260 nm circularly polarized laser pulse at the appropriate position in the wake bubble where the longitudinal and transverse electric fields are zero (Fig.\,\ref{fig2}(a)), electrons with a net spin polarization are generated and injected into the wakefield. The trapping condition is given by \cite{pak2010injection} $\Delta \Psi \equiv \Psi-\Psi_\text{init}\lesssim -1$, where $\Psi\equiv \frac{e(\phi-A_z )}{mc^2}$ is the normalized pseudo potential of the wake, and $\Psi_\text{init}$ is the pseudo potential at the position where the electron is born (injected). The pseudo potential is maximum at the center of the bubble and minimum close to the rear. For this reason, we choose to inject electrons where $\Psi_\text{init}$ is maximum so that the injected electrons are most easily trapped by the wake (Fig.\,\ref{fig2}(a,b)).

Previous studies have shown that spin dynamics due to the Stern-Gerlach force, the Sokolov-Ternov effect (spin flip), and radiation reaction force are negligible in our case \cite{vieira2011polarized,wu2019polarized,wen2019polarized,wu2019polarized1}. Therefore, only spin precession needs to be considered. We have implemented the spin precession module into the 3D-PIC code OSIRIS \cite{fonseca2002osiris,fonseca2008one} following the Thomas-Bargmann-Michel-Telegdi (T-BMT) equation  \cite{bargmann1959precession}
\begin{align}\label{T-BMT}
d\bold{s}/dt=\bold{\Omega}\times \bold{s}
\end{align}
where $\bold{\Omega}=\frac{e}{m} (\frac{1}{\gamma} \bold{B}-\frac{1}{\gamma+1} \frac{\bold{v}}{c^2} \times \bold{E})+a_e\frac{e}{m} [\bold{B}-\frac{\gamma}{\gamma+1} \frac{\bold{v}}{c^2} (\bold{v}\cdot \bold{B})-\frac{\bold{v}}{c^2} \times \bold{E}]$ . Here, $\bold{E},\bold{B}$ are the electric and magnetic field, $\bold{v}$ is the electron velocity, $\gamma=\frac{1}{\sqrt{1-v^2/c^2} }$ is the relativistic factor, and $a_e\approx 1.16\times 10^{-3}$ is the anomalous magnetic moment of the electron. 

As shown in Fig.\,\ref{fig2}(c) and (d), the spin vector distribution is at first concentrated around the top and bottom points of $s_z=\pm1$ with a very small spread when the Xe electrons are photoionized (Fig.\,\ref{fig2}(c)), caused by the spread of the ionizing laser wavevectors at different ionization positions. In our case, this initial spread of the spin vector is within $1^{\circ}$, which is negligible compared to the spread due to spin precession induced by the wakefield at later times (Fig.\,\ref{fig2}(d)).

\begin{figure}[tp]
\includegraphics[width=0.5\textwidth]{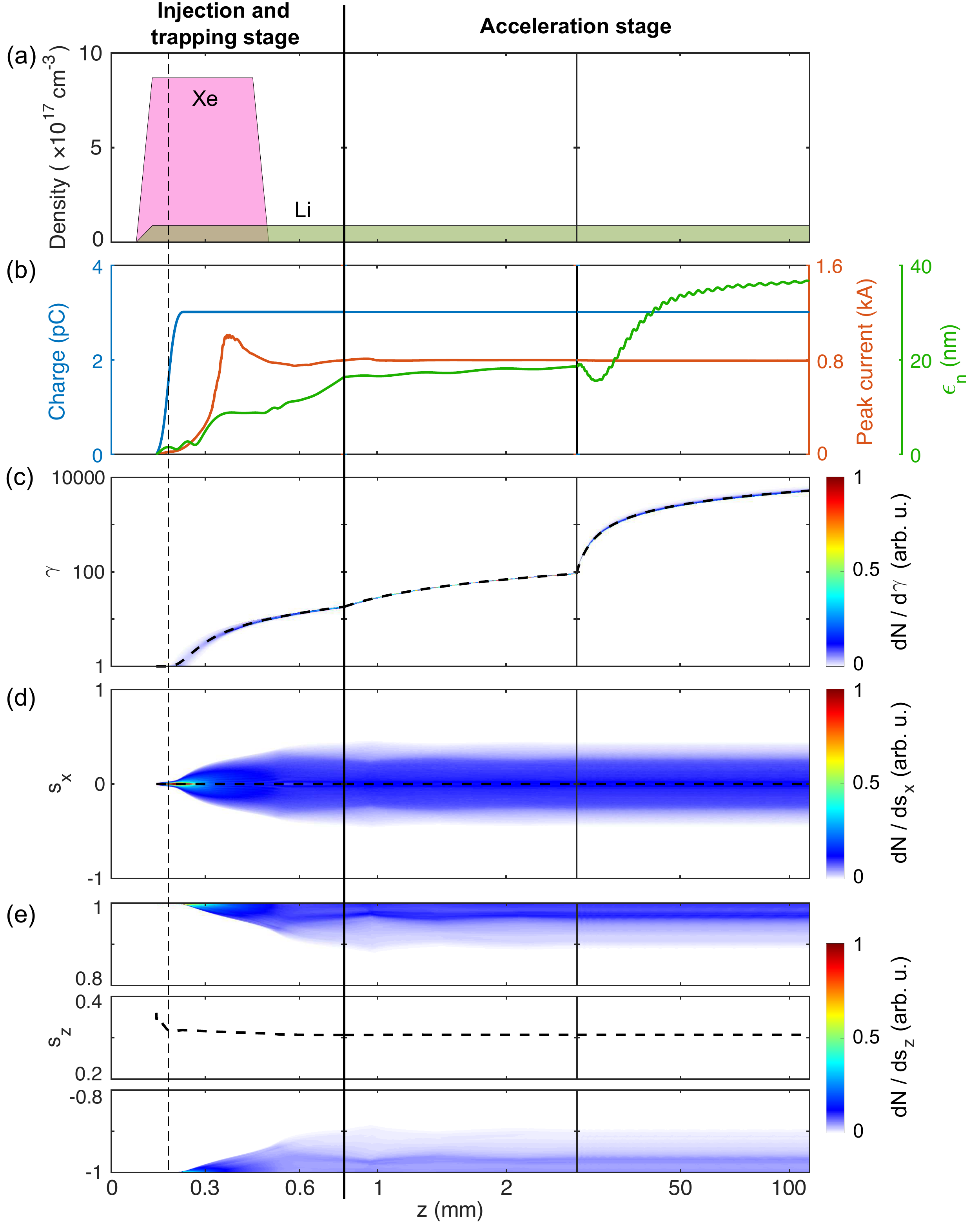}
\caption{\label{fig3} 
(a) The density profiles of the Xe and Li gases used in the simulations. (b) Evolution of beam charge (left blue axis), peak current (right red axis) and normalized emittance $\epsilon_n$ (right green axis). (c) Evolution of Lorentz factor $\gamma$. The dashed line presents mean energy $\langle \gamma\rangle$. d, Evolution of spin vector in the x direction: $s_x$. The dashed line represents $\langle s_x\rangle$. e, Evolution of spin vector in the z direction: $s_z$. The top box plots the $s_z$ distribution in the range of 0.8 and 1. The central box plots $\langle s_z\rangle$ (net spin polarization) in the range of 0.2 and 0.4. The bottom box plots the $s_z$ distribution in the range of $-1$ and $-0.8$. The long vertical dashed black line marks the focal position ($z=0.18$\,mm) of the ionization laser. The plots in the range of 0.74-2.5\,mm and 2.5-110\,mm are shown in two temporal scales to clearly present the whole evolution dynamics but the actual simulation was run with one temporal resolution in the whole acceleration stage.
}
\end{figure}

Figure \ref{fig3} describes start-to-end simulations incorporating both the TDSE and PIC components. The whole simulation consists of two stages: the injection and trapping stage (0-0.74 mm) and acceleration stage (0.74-110 mm). The injection and trapping stage was simulated using the OSIRIS code \cite{fonseca2002osiris,fonseca2008one} with high temporal resolution and the acceleration stage was simulated using the QPAD code  \cite{li2020quasi,sprangle1990nonlinear}  with lower temporal resolution. The density profiles of Xe and Li gases are shown in Fig.\,\ref{fig3}(a). The Xe gas column, with a density of $n_\text{Xe}=8.7\times 10^{17}\,\text{cm}^{-3}$, is 420\,$\mu$m long. The exact length of the Xe region is not important as long as Xe is not ionized by the electron beam. The Li gas, with a density of $n_\text{Li}=8.7\times 10^{16}\,\text{cm}^{-3}$, extends across the whole interaction region and provides background plasma electrons when ionized by the drive electron beam. The driving beam electron energy is 10\,GeV with a Gaussian profile $n_b=\frac{N}{(2\pi)^{3/2} \sigma_r^2 \sigma_z}\,\text{exp}(-\frac{r^2}{2 \sigma_r^2}-\frac{\xi^2}{2 \sigma_z^2})$, where $N=4.11 \times 10^9$ (658\,pC), and $\sigma_r=\sigma_z=11.4\,\mu$m are the transverse and longitudinal beam sizes, respectively. Such a beam has a maximum electric field of 16\,GV/m, which is far larger than that required to fully ionize the Li atoms, but not the Xe atoms. It forms the plasma and blows out the plasma electrons to create the wake cavity. The 260\,nm ionization laser is delayed by 148\,fs (44.5\,$\mu$m) from the peak current position of the drive electron beam. The laser pulse has Gaussian envelope with pulse duration (FWHM) of 30\,fs and focal spot size of $w_0 = 1.5\,\mu$m. The peak laser intensity is $2.5\times 10^{13}\,\text{W/cm}^2$ (the same intensity as in Fig.\,\ref{fig1}(b)) to make a tradeoff between net spin polarization and ionization yield. At this peak laser intensity, the $5p^6$ (outermost) electron of Xe is partially ionized ($\sim32\%$ at focus) while the $5p^5$ (second) electron of Xe is not ionized at all ($<10^{-6}$).

Evolution of injected beam parameters including charge, peak current, normalized emittance, and spin vector distribution as a function of propagation distance in the plasma are shown in Fig.\,\ref{fig3}(b)-(e). All photoionized electrons with charge of 3\,pC (Fig.\,\ref{fig3}(b) left axis) are injected, trapped and accelerated to 2.7\,GeV (Fig.\,\ref{fig3}(c)) within 11\,cm to give a peak current of $I=0.8$\,kA (Fig.\,\ref{fig3}(b) right red axis) and normalized transverse emittance of $\epsilon_n=36.6$\,nm (Fig.\,\ref{fig3}(b) right green axis). This emittance compares favorably with the brightest beams available today \cite{schmerge2015lcls}. The spin vector evolutions in the $x$ and $z$ directions are shown in Fig.\,\ref{fig3}(d) and (e), respectively. The spin spread in $x$ (or $y$) direction is symmetric so that $\langle s_x\rangle\approx 0$ (or $\langle s_y\rangle\approx 0$) as shown in Fig.\,\ref{fig3}(d). Therefore, the net spin polarization $P=P_z=\langle s_z\rangle$ only depends on the spin distribution in the $z$ direction. The spin depolarization mainly occurs during the first 500\,$\mu$m distance as electrons are injected into the wake until they become ultra-relativistic ($\gamma\sim 10$). Thereafter the spin polarization remains constant within the statistical sampling error. The final averaged spin polarization is $\langle s_z\rangle=30.7\%$ (Fig.\,\ref{fig3}(e)), corresponding to 96\% of the initial spin polarization at birth. This result is comparable to the first-generation GaAs polarized electron sources, that are most commonly used in conventional rf accelerators. The reason why depolarization is small in our case is that the injected electrons are always close to the axis of the wake so that the transverse magnetic and electric fields they feel are close to zero. In a nonlinear wake bucket, the transverse magnetic field $B_\phi$ scales linearly with distance from the center of the wake ($B_\phi\propto r$) \cite{lu2006nonlinear}. From Eq.\,(\ref{T-BMT}), the spin precession frequency $\Omega\approx -eB_\phi/m\gamma$ when $\gamma\sim 1$. Therefore, if the electrons are close to the axis ($r\approx 0$), the spin precession frequency $\Omega\approx 0$. In addition, once the electron energy is increased to ultra-relativistic level ($\gamma\gg 1$) by the longitudinal wakefield, the spin precession effect is negligible \cite{vieira2011polarized}.

\begin{figure}[tp]
\includegraphics[width=0.5\textwidth]{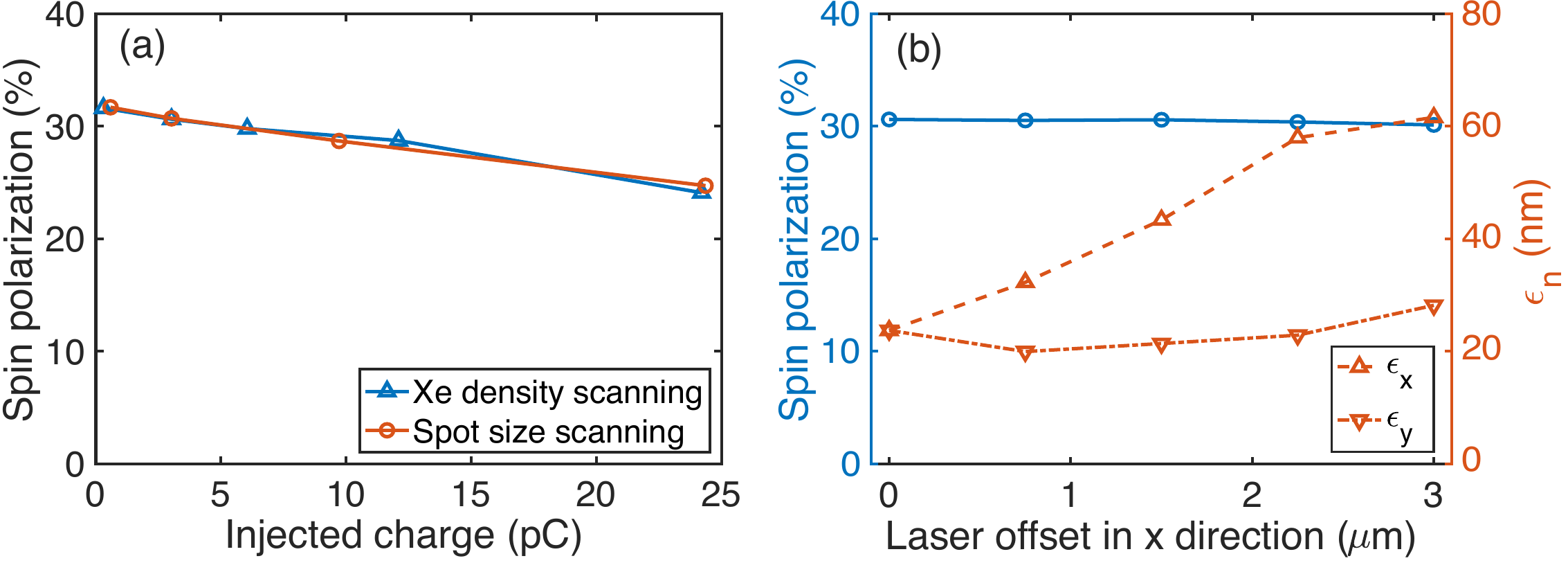}
\caption{\label{fig4} 
(a) Spin polarization v.s. injected beam charge by either varying the Xe density (blue) or the spot size of the ionization laser (red). The five data points of Xe density scanning correspond to Xe density of $8.7\times 10^{16}\,\text{cm}^{-3}$, $8.7\times 10^{17}\,\text{cm}^{-3}$, $1.7\times 10^{18}\,\text{cm}^{-3}$, $3.5\times 10^{18}\,\text{cm}^{-3}$, and $7.0\times 10^{18}\,\text{cm}^{-3}$ while keeping the spot size of 1.5\,$\mu$m. The four data points of spot size scanning correspond to ionization laser spot size of 1\,$\mu$m, 1.5\,$\mu$m, 2\,$\mu$m, and 2.5\,$\mu$m while keeping the Xe density of $8.7\times 10^{17}\,\text{cm}^{-3}$. (b) Spin polarization (left) and normalized emittance (right) after propagation distance of 0.74 mm v.s. laser transverse displacement in x direction.
}
\end{figure}

We have investigated how the variation of injected beam charge (by either varying the Xe density or the spot size of the ionization laser) affects the final spin polarization of the injected electrons. The parameter scanning results are summarized in Fig.\,\ref{fig4}(a). The spin polarization drops slowly and linearly with the increase of the beam charge. This indicates that the space charge force is the probable cause of spin depolarization in our case, which is confirmed by analyzing the tracks of the ionized electrons (see Supplementary Material \cite{SM} for details). Considering practical issues in experiments, we have investigated how the laser transverse offset relative to the drive electron beam affects the spin polarization and normalized emittance as shown in Fig.\,\ref{fig4}(b). The spin polarization is essentially not affected by the transverse displacement in $\pm3\,\mu$m range. The normalized emittance in $x$ direction grows with the laser offset in $x$ direction and the normalized emittance in $y$ direction remains almost the same. These emittances are within values envisioned for future plasma-based colliders. Another possible issue in experiments might be the synchronization between the drive electron beam and the ionizing laser pulse. To make sure the ionized electrons are trapped by the wake (meet the trapping condition $\Delta \Phi \lesssim -1$), the relative timing jitter should be within $\pm 80$\,fs in our simulation case. This requirement can be further relaxed if using higher drive beam charge and lower plasma density.

Here we have used a single collinearly (to the electron beam) propagating laser pulse for ionizing the Xe atoms. To obtain even lower emittance ($<$10\,nm) beams, one could use two transverse \cite{li2013generating} or longitudinal \cite{hidding2012ultracold} colliding laser pulses instead. We also note that the beam charge, peak current, and the maximum spin polarization observed here are not limited by theory. The first two can be increased by optimizing the ionizing laser parameters, drive beam parameters, and the beam loading within the wake. The latter may be increased by using electrons in the $d$ or $f$ orbitals instead of $p$ orbitals – for instance by using Yb III \cite{kaushal2018looking,kaushal2018looking1}. A modified version of this scheme may also be useful for generating a spin-polarized electron beam in a laser wakefield accelerator (LWFA) \cite{xu2014low}. 

\begin{acknowledgments}
We thank Nuno Lemos and Christopher E. Clayton for useful discussions regarding this work. This work was supported by AFOSR grant FA9550-16-1-0139, DOE grant DE-SC0010064, DOE SciDAC through FNAL subcontract 644405, and NSF grants 1734315 and 1806046. The simulations were performed on Hoffman cluster at UCLA and NERSC at LBNL.
\end{acknowledgments}



\end{document}